\documentclass[twocolumn,aps,nofootinbib,superscriptaddress]{revtex4}

\usepackage{amsmath}
\usepackage{amsthm}
\usepackage{amsfonts}
\usepackage{color}
\usepackage{mathtools}
\usepackage{braket}

\usepackage[ colorlinks = true,
             linkcolor = blue,
             urlcolor  = blue,
             citecolor = red,
             anchorcolor = green,
]{hyperref}

\newtheorem{lemma}{Lemma}

\newtheorem{theorem}[lemma]{Theorem}

\DeclareMathOperator{\tr}{tr}

\newcommand{\eps}{\varepsilon}
\newcommand{\ketbra}[1]{\ket #1\!\!\bra #1}

\begin{document}

\title{Disentanglement Cost of Quantum States}

\author{Mario Berta}
\affiliation{Department of Computing, Imperial College London, London}

\author{Christian Majenz}
\affiliation{Institute for Language, Logic and Computation, University of Amsterdam, Amsterdam}
\affiliation{QuSoft, Amsterdam}

\begin{abstract}
We show that the minimal rate of noise needed to catalytically erase the entanglement in a bipartite quantum state is given by the regularized relative entropy of entanglement. This offers a solution to the central open question raised in [Groisman {\it et al.}, PRA 72, 032317 (2005)] and complements their main result that the minimal rate of noise needed to erase all correlations is given by the quantum mutual information. We extend our discussion to the tripartite setting where we show that an asymptotic rate of noise given by the regularized relative entropy of recovery is sufficient to catalytically transform the state to a locally recoverable version of the state.
\end{abstract}

\maketitle


\paragraph*{Introduction}

Quantifying and classifying quantum correlations is of fundamental importance in quantum information theory~\cite{Horodecki09}. Motivated by Landauer's erasure principle~\cite{landauer1961irreversibility} one way to quantify the correlations present in a bipartite quantum state $\rho_{AB}$ is to measure the amount of noise that is required to erase them. In that respect, Groisman {\it et al.}~\cite{groisman2005quantum} showed that the optimal asymptotic rate of local noise to bring $\rho_{AB}$ close to a product $\sigma_A\otimes\sigma_B$ is given by the {\it quantum mutual information}
\begin{align}\label{eq:quantum_mutual}
I(A:B)_\rho:=D(\rho_{AB}\|\rho_A\otimes\rho_B)=\inf_{\sigma\in\mathrm{PR}}D(\rho_{AB}\|\sigma_A\otimes\sigma_B)
\end{align}
with $\text{PR}(A:B)$ the set of product states in $A:B$ and $D(\rho\|\sigma):=\mathrm{Tr}\left[\rho(\log\rho-\log\sigma\right)]$ is the quantum relative entropy. Hence, the quantum mutual information quantifies the total amount of correlations in bipartite states -- including both the quantum and classical ones. Alternatively, we can write
\begin{align}\label{eq:quantum_mutual2}
I(A:B)_\rho=\inf_{\sigma\in\mathrm{PR}}H(AB)_{\sigma\otimes\sigma}-H(AB)_\rho
\end{align}
with $H(A)_\rho:=-\tr\left[\rho_A\log\rho_A\right]$ the von Neumann entropy. Thus, the cost function $I(A:B)_\rho$ can conveniently be understood as either the quantum relative entropy distance to the next product state as in Eq.~\eqref{eq:quantum_mutual} or as the amount of entropy injected into the system to reach the next product state as in Eq.~\eqref{eq:quantum_mutual2}. This finding was generalized in various directions, including a catalytic analysis of the one-shot case~\cite{Majenz17}, the study of tripartite correlations~\cite{Berta2018,Wakakuwa17,Wakakuwa17_2}, as well as the study of coherence~\cite{Singh15} and more general symmetries~\cite{Wakakuwa17_3}. However, it remained open how to quantify the optimal asymptotic rate of local noise to bring $\rho_{AB}$ close to a separable state $\sum_jp_j\sigma^j_A\otimes\sigma^j_B$. In particular, it was unclear if a quantity defined in such a way can be the basis of a proper entanglement measure. 

In this note, we solve the problem and give a precise mathematical model for erasing entanglement in bipartite states where the optimal asymptotic rate of local noise needed to get close to a separable state is given by the {\it regularized relative entropy of entanglement}. In particular, this also gives a new operational interpretation to the distance measure quantum relative entropy.


\paragraph*{Entanglement Measures.}

As discussed the quantum mutual information is a measure for the total amount of correlations and in the following we introduce more refined measures only capturing the quantum correlations. The {\it relative entropy of entanglement} is given by~\cite{Vedral97}
\begin{align}
E(A:B)_\rho:=\inf_{\sigma\in\mathrm{SEP}}D(\rho\|\sigma)\,,
\end{align}
where $\mathrm{SEP}(A:B)$ denotes the set of separable states in $A:B$. Since the relative entropy of entanglement is in general not additive
on tensor product states, it has to be regularized \cite{vollbrecht01,Audenaert2001}. The
regularized relative entropy of entanglement is defined as
\begin{align}
E^\infty(A:B)_\rho:=\lim_{n\to\infty}\frac{1}{n}E(A:B)_{\rho^{\otimes n}}
\end{align}
This quantity has an operational interpretation in composite asymmetric quantum hypothesis testing as the asymptotic exponential rate of mistakenly identifying $\rho_{AB}$ instead of a state separable in $A:B$~\cite{brandao10}. As a corresponding one-shot analogue based on the smooth max-relative entropy~\cite{datta08}
\begin{align}
&D_{\max}^{\eps}(\rho\|\sigma):=\inf_{\bar{\rho}\approx_{\eps}\rho}\inf\left\{\lambda:\;2^\lambda\cdot\sigma-\bar{\rho}\geq0\right\}\\
&\text{with $\bar{\rho}\approx_{\eps}\rho$ in purified distance~\cite{tomamichel09},}\notag
\end{align}
we have the {\it smooth max-relative entropy of entanglement}~\cite{datta09_2}
\begin{align}
&E_{\max}^{\eps}(A:B)_\rho:=\inf_{\sigma\in\mathrm{SEP}}D_{\max}^{\eps}(\bar{\rho}_{AB}\|\sigma_{AB})\,.
\end{align}
This is a smoothed version of the logarithm global robustness of entanglement~\cite{Vidal99}. All the quantities
\begin{align}
\text{$E(A:B)_\rho$, $E^\infty(A:B)_\rho$, $E_{\max}^{\eps}(A:B)_\rho$}\notag
\end{align}
define proper entanglement measures with mathematical properties as requested by axiomatic entanglement theory (see, e.g., \cite{matthiasphd,brandao11} for an overview). We emphasize that these type of information-theoretic entanglement measures have been vastly useful for understanding the entanglement structure of multipartite quantum states in many body physics. In particular, this lead to strong insights about entropic area laws~\cite{Hastings07,Wolf08,horodecki13,brandao_horodecki15} and detecting topological order~\cite{PK06,LW06} in condensed matter physics as well as to basics findings in quantum thermodynamics~\cite{delrio11,groisman2005quantum,plenio10,Gour15}.


\paragraph*{Disentanglement Cost.}

We are interested in the amount of local noise needed to catalytically erase the entanglement in a bipartite quantum state. For this purpose -- following Groisman {\it et al.}~\cite{groisman2005quantum} and the follow-up works~\cite{Singh15,Berta2018,Wakakuwa17,Wakakuwa17_2,Majenz17,Wakakuwa17_3} -- a randomizing map is generated by an ensemble of local unitaries $(U_A^i\otimes U_B^i)$ as
\begin{align}\label{eq:map}
\Lambda^M_{A:B}(\cdot):=\frac{1}{M}\sum_{i=1}^M\left(U_A^i\otimes U_B^i\right)(\cdot)\left(U_A^i\otimes U_B^i\right)^\dagger\,.
\end{align}
It is called {\it $\eps$-disentangling} if there exist a state $\omega_{A'B'}\in\mathrm{SEP}(A':B')$ such that 
\begin{align}\label{eq:goal}
\inf_{\sigma\in\mathrm{SEP}}P\left(\Lambda^M_{AA':BB'}(\rho_{AB}\otimes\omega_{A'B'}),\sigma_{ABA'B'}\right)\leq\eps
\end{align}
with $\sigma_{AA'BB'}\in\mathrm{SEP}(AA':BB')$. Here we think of $\omega_{A'B'}$ as a catalytic resource state that is already separable to start with but has to be kept separable by the randomizing map (cf.~catalytic decoupling~\cite{Majenz17}). The {\it one-shot $\eps$-disentanglement cost} $C^\eps_{\mathrm{SEP}}(A:B)_\rho$ is then defined as the minimal number $\log M$ such that Eq.~\eqref{eq:goal} holds. We are particularly interested in the asymptotic behavior in the limit of many copies $\rho_{AB}^{\otimes n}$ and vanishing error $\eps\to0$, which we call the {\it disentanglement cost} of quantum states:
\begin{align}
C_{\mathrm{SEP}}(A:B)_\rho:=\lim_{\eps\to0}\lim_{n\to\infty}\frac{1}{n}C^\eps_{\mathrm{SEP}}(A:B)_{\rho^{\otimes n}}\,.
\end{align}


\paragraph*{Main Result.}

We find that the $\eps$-disentanglement cost is given by the smooth max-relative entropy of entanglement and hence that the disentanglement cost is given by the regularized relative entropy of entanglement.

\begin{theorem}\label{thm:distangle}
Let $\rho_{AB}$ and $1\geq\eps\geq\delta>0$. Then, we have
\begin{align}
E_{\max}^{\eps}(A:B)_\rho&\leq C^\eps_{\mathrm{SEP}}(A:B)_\rho\label{eq:distangle_1}\\
&\leq E_{\max}^{\eps-\delta}(A:B)_\rho+\log\frac{1}{\delta}\label{eq:distangle_2}
\end{align}
as well as $C_{\mathrm{SEP}}(A:B)_\rho=E^{\infty}(A:B)_\rho$.
\end{theorem}

This offers a solution to the central open question raised in Groisman {\it et al.}~\cite{groisman2005quantum} and automatically establishes the disentanglement cost of quantum states as a proper {\it entanglement measure} -- since it inherits all mathematical properties from the regularized relative entropy of entanglement. Note, however, that we do not show the disentanglement cost being equal to the asymptotic rate of entropy injected into the system as conjectured by Groisman {\it et al.} (cf.~Eq.~\eqref{eq:quantum_mutual})
\begin{align}
\frac{1}{n}\inf_{\sigma\in\mathrm{SEP}}\Big(H(AB)_{\sigma_n}-H(AB)_{\rho^{\otimes n}}\Big)\,,
\end{align}
but to the relative entropy of entanglement as suggested in~\cite{Horodecki05_2}. For pure states $\ket{\psi}_{AB}$ we get $E^{\infty}(A:B)_\psi=H(A)_\psi$ -- the entropy of the Schmidt spectrum -- whereas the quantum mutual information measuring the total correlations is equal to $2H(A)_\psi$. For the one-shot setting we find that
\begin{align}\label{eq:pure_state}
&H_{\max}^{\eps}(A)_\psi\leq C^\varepsilon_{\mathrm{SEP}}(A:B)_\psi \leq H_{\max}^{\eps-\delta}(A)_\rho+\log\frac{1}{\delta}\\
&\text{with $H_{\max}^{\eps}(A)_\rho:=\inf_{\bar{\rho}\approx_{\eps}\rho}2\log\mathrm{Tr}\left[\sqrt{\bar{\rho}}\right]$}\notag
\end{align}
the smooth max-entropy. Furthermore, we find with~\cite{Majenz17} that the amount of noise needed to erase all correlations in a pure state $\ket{\psi}_{AB}$ is given by two times the cost function from Eq.~\eqref{eq:pure_state} -- which is in exact analogy to the asymptotic case.


\paragraph*{Proof of Thm.~\ref{thm:distangle}.}

We first derive the converse direction -- i.e.~the lower bound in Thm.~\ref{thm:distangle} -- using standard entropy inequalities. To show the one-shot converse in Eq.~\eqref{eq:distangle_1} we begin by observing that tensoring a separable state does not change the smooth max-relative entropy of entanglement\footnote{The argument is the same as for the relative entropy of entanglement~\cite{vollbrecht01} and based on the monotonicity under quantum operations.} and thus it suffices to show the converse for disentangling maps without catalysts. Let therefore $\Lambda^M_{A:B}$ be a disentangling randomizing map for $\rho_{AB}$, that is, there exists $\sigma_{AB}\in\mathrm{SEP}(A:B)$ such that $P\left(\Lambda^M_{A:B}(\rho_{AB}),\sigma_{AB}\right)\leq\eps$. Next, define a classically maximally correlated state 
\begin{equation}\label{eq:gamma}
\gamma_{X_aX_b}:=\frac{1}{M}\sum_{i=1}^M\ketbra{i}_{X_a}\otimes \ketbra{i}_{X_b}
\end{equation}
and the controlled unitaries $V_{AX_a}$ and $W_{BX_b}$ such that
\begin{align}
\tr_{X_aX_b}\left[\rho'_{ABX_aX_b}\right]=\Lambda^M_{A:B}(\rho_{AB})
\end{align}
for the state
\begin{align}
&\rho'_{ABX_aX_b}\notag\\
&:=\left(V_{AX_a}\otimes W_{BX_b}\right)\left(\rho_{AB}\otimes\gamma_{X_aX_b}\right)\left(V_{AX_a}\otimes W_{BX_b}\right)^\dagger\,.
\end{align}
By Uhlmann's theorem, there exists an extension $\sigma_{ABX_aX_b}$ of $\sigma_{AB}$ such that $P\left(\rho'_{ABX_aX_b},\sigma_{ABX_aX_b}\right)\leq\eps$ with the $X_a$- and $X_b$-registers classical in the same basis as in Eq.~\eqref{eq:gamma}. Additionally, the extension can be chosen such that $\Pi_{X_aX_b}\sigma_{ABX_aX_b}\Pi_{X_aX_b}=\sigma_{ABX_aX_b}$, where $\Pi_{X_aX_b}$ is the projector onto the maximally correlated subspace, i.e. onto the support of $\gamma_{X_aX_b}$. Now we bound
\begin{align}
E_{\max}^\varepsilon(A:B)_\rho=&\;E_{\max}^\varepsilon(AX_a:BX_b)_{\rho\otimes\gamma}\nonumber\\
=&\;E_{\max}^\varepsilon(AX_a:BX_b)_{\rho'}\nonumber\\
\le&\;E_{\max}^0(AX_a:BX_b)_{\sigma}\nonumber\\
\le&\;D_{\max}^0(\sigma_{ABX_aX_b}\|\sigma_{AB}\otimes\gamma_{X_aX_b})\nonumber\\
\le&\;\log M\,.
\end{align}
The first two inequalities follow from picking two particular points in the minima defining $E_{\max}^\varepsilon$ and the last inequality follows from the matrix inequality 
\begin{align}\label{eq:matrixineq}
\sigma_{ABX_aX_b}\leq&\;\sigma_{AB}\otimes\Pi_{X_aX_b}=M\cdot\sigma_{AB}\otimes\gamma_{X_aX_b},
\end{align} 
which is as in~\cite[Lem.~3.1.9]{renner05}. This proves Eq.~\eqref{eq:distangle_1}.

For the asymptotic expansion, we then use the composite quantum Stein's lemma from~\cite[Prop.~II.1]{brandao10} and~\cite[Thm.~1]{datta09_2}
\begin{align}\label{eq:asymptotic_expansion}
\lim_{\eps\to0}\lim_{n\to\infty}\frac{1}{n}E_{\max}^{\eps}(A\!:\!B)_\rho=E^{\infty}(A\!:\!B)_\rho\,.
\end{align}
We note that asymptotic converses for similar scenarios were also shown in~\cite{Horodecki05_2}.

For the achievability part -- i.e.~the upper bound in Thm.~\ref{thm:distangle} -- we invoke a tool that was introduced as the {\it convex splitting lemma} by Anshu {\it et al.}~\cite{Anshu14}. We need a special case of their main lemma which is as follows.

\begin{lemma}[Convex split]\label{lem:convex_split}
Let $\rho,\sigma$ be quantum states and $N=\left\lceil 2^{D^{\zeta}_{\max}(\rho\|\sigma)}/\xi\right\rceil$ with $\xi>0,\zeta\geq 0$. Then, we have
\begin{align}
P\left(\frac{1}{N}\sum_{i=1}^N\rho_i\otimes\sigma^{\otimes(N-1)}_{i^c},\sigma^{\otimes N}\right)\leq\zeta+\xi\,,
\end{align}
where $\rho_i$ sits in the $i$-th register and $i^c:=[1,\ldots,N]\backslash i$.
\end{lemma}

We emphasize that this convex split lemma is neatly proven only using elementary properties of quantum entropy~\cite[Lem.~12]{Anshu14}. Now for any state $\rho_{AB}$ and $\sigma_{AB}\in\mathrm{SEP}(A:B)$ we can choose $\log N=D_{\max}^{\eps-\delta}(\rho_{AB}\|\sigma_{AB})+\log\frac{1}{\delta}$ in Lem.~\ref{lem:convex_split} such that
\begin{align}
P\left(\frac{1}{M}\sum_{i=1}^M\rho_{A_iB_i}\otimes\sigma^{\otimes(M-1)}_{\tilde{A}\tilde{B}\backslash(A_iB_i)},\sigma^{\otimes M}_{\tilde{A}\tilde{B}}\right)\leq\eps
\end{align}
for $\tilde{A}\tilde{B}:=A_1\cdots A_MB_1\cdots B_M$ with $A_1B_1:=AB$ and $A_iB_i\cong AB$ for $i=2,\ldots,M$. The idea is to use the catalytic resource state $\sigma_{\tilde{A}\tilde{B}\backslash(A_1B_1)}^{\otimes(M-1)}\in\mathrm{SEP}(\tilde{A}\backslash A_1:\tilde{B}\backslash B_1)$ together with the ensemble of local unitaries for $i=1,\ldots,N$ given by
\begin{align}
U_{\tilde{A}}^i\otimes U_{\tilde{B}}^i:=(1i)_{\tilde{A}}\otimes(1i)_{\tilde{B}}\,,
\end{align}
where $(1i)$ denotes the unitary that swaps registers $1\leftrightarrow i$ on $\tilde{A}$ and $\tilde{B}$, respectively. Optimizing over all $\sigma_{AB}\in\mathrm{SEP}(A:B)$ then gives the one-shot achievability in Eq.~\eqref{eq:distangle_2}. Finally, the asymptotic expansion of the upper bound follows as in Eq.~\eqref{eq:asymptotic_expansion} which concludes the proof of Thm~\ref{thm:distangle}. \qed


\paragraph*{Multipartite Extension.}

The relative entropy of entanglement can naturally be extended to the multi-party setting (see, e.g., \cite{plenio2005introduction}). For a $k$-party quantum state $\rho_{A_1,...,A_k}$ it is defined as the relative entropy distance to the set $\mathrm{SEP}$ of completely separable states, 
\begin{align}
E(A_1\!:\!A_2:...\!:\!A_k)_\rho:=\inf_{\sigma\in\mathrm{SEP}}D(\rho\|\sigma)\,.
\end{align}
A regularized version $E^\infty(A_1\!:\!A_2\!:...:\!A_k)_\rho$ is defined the same way as in the two-party setting. It is then straightforward to generalize our Theorem~\ref{thm:distangle} to the multi-party setting: $E^\infty(A_1\!:\!A_2\!:...:\!A_k)_\rho$ is equal to the \emph{multiparty disentanglement cost}, i.e. the asymptotic noise rate that is necessary to transform $\rho_{A_1...A_k}^{\otimes n}$ into a fully separable state, for $n\to\infty$.


\paragraph*{Catalytic Decoupling.}

Groisman {\it et al.}~\cite{groisman2005quantum} show that for their setting of going to product states one can also achieve the quantum mutual information by alternatively replacing the model of coordinated random local unitary channels as in Eq.~\eqref{eq:map} to only local unitary channels $\Lambda_A^M(\cdot):=\frac{1}{M}\sum_{i=1}^MU_A^i(\cdot)\left(U_A^i\right)^\dagger$ and not making use of any (product state) catalytic assistance. Whereas maps as in Eq.~\eqref{eq:map} and catalytic assistance -- separable states in our case -- seem necessary to obtain the tight result presented in the previous sections, it is nevertheless insightful to compare our result with other models. In particular, the model of local unitary channels $\Lambda_A^M(\cdot)$ can be related to {\it catalytic decoupling}, where the noisy operation to ensure closeness to product states is given by a {\it partial trace map} over a system of asymptotic rate size $\frac{1}{2}I(A:B)_\rho$~\cite{Majenz17}. This can be done in our case as well, albeit not in the exact same optimal way as for local unitary channels. Namely, to implement the coordinated local random unitary channel from Eq.~\eqref{eq:map}, a classically correlated state $\gamma_{X_aX_b}$ has to be used as an ancillary system, half of which has to be discarded afterwards on both sites $A$ and $B$. More precisely for
\begin{align}
\mu_{\bar{A}\bar{B}}:=\rho_{AB}\otimes\omega_{A'B'}\otimes\gamma_{X_aX_b}&\\
\text{with $\omega_{A'B'}\in\mathrm{SEP}(A':B')$}&\notag
\end{align}
and $\bar{A}\bar{B}:=\bar{A}_1\bar{A}_2\bar{B}_1\bar{B}_2:=AA'X_aBB'X_b$ there exist $\sigma_{\bar{A_1}\bar{B}_1}\in\mathrm{SEP}(\bar{A}_1:\bar{B}_1)$ and a local unitary $U_{\bar{A}}\otimes U_{\bar{B}}$ such that
\begin{align}\label{eq:separable_decoupling}
P\left(\mathrm{Tr}_{\bar{A}_2\bar{B}_2}\left[\left(U_{\bar{A}}\otimes U_{\bar{B}}\right)\mu_{\bar{A}\bar{B}}\left(U_{\bar{A}}\otimes U_{\bar{B}}\right)^\dagger\right],\sigma_{\bar{A_1}\bar{B}_1}\right)\leq\eps
\end{align}
for $\log|\bar{A}_2|+\log|\bar{B}_2|=E_{\max}^{\eps-\delta}(A:B)_\rho+\log\frac{1}{\delta}+1$. We conclude that the straightforward translation of the disentangling protocol introduced here to two-sided catalytic decoupling leads to a cost twice the one obtained from the converse bound in the case of disentangling. It would be interesting to explore the decoupling to separable states notion as in Eq.~\eqref{eq:separable_decoupling} further.


\paragraph*{Tripartite Correlations.}

We might extend our results to analyze tripartite quantum correlations as well. Here for tripartite states $\rho_{ABC}$ we can define {\it locally recovered states} by
\begin{align}
&(\mathcal{I}_B\otimes\mathcal{R}_{C\to AC})(\rho_{BC})\\
&\text{with $\mathcal{R}_{C\to AC}$ local quantum channels.}\notag
\end{align}
States $\rho_{ABC}$ such that there exists $\mathcal{R}_{C\to BC}$ with $(\mathcal{I}_B\otimes\mathcal{R}_{C\to AC})(\rho_{BC})=\rho_{ABC}$ are called {\it quantum Markov}~\cite{petz86b} but in general $\rho_{ABC}$ is far from its recovered states. A measure for the local recoverability is the relative entropy of recovery
\begin{align}
&D(A;B|C)_\rho\nonumber\\
&:=\inf_{\mathcal{R}_{C\to BC}}D\big(\rho_{ABC}\|(\mathcal{I}_B\otimes\mathcal{R}_{C\to AC})(\rho_{BC})\big)
\end{align}
and its regularized version $D^{\infty}(A;B|C)_\rho$~\cite{Brandao15,Seshadreesan15}. The latter quantity has an operational interpretation in composite asymmetric quantum hypothesis testing as the asymptotic exponential rate of mistakenly identifying $\rho_{ABC}$ instead of a corresponding locally recovered state $(\mathcal{I}_B\otimes\mathcal{R}_{C\to BC})(\rho_{AC})$~\cite{cooney16}. Moreover, it was recently shown that~\cite{Fawzi16}
\begin{align}
D^{\infty}(A;B|C)_\rho\neq D(A;B|C)_\rho\,.
\end{align}
We can now ask for the amount of noise needed to catalytically transform the state into a corresponding locally recovered version thereof. For this purpose we again define a randomizing map $\Lambda_{ABC}^M$ as in Eq.~\eqref{eq:map} but now with tripartite local unitaries $\left(U_A^i\otimes U_B^i\otimes U_C^i\right)$. Such maps are called {\it $\eps$-recovery-degrading} if there exists a locally recovered state $\omega_{A'B'C'}=(\mathcal{I}_{B'}\otimes\mathcal{R}_{C'\to A'C'})\left(\rho_{BC}^{\otimes(M-1)}\right)$ such that
\begin{align}
&\inf_{\mathcal{R}_{CC'\to AA'CC'}}P\big(\Lambda^M_{AA'BB'CC'}(\rho_{ABC}\otimes\omega_{ABC}^{\otimes {(M-1)}}),\notag\\
&(\mathcal{I}_{BB'}\otimes\mathcal{R}_{CC'\to AA'CC'})(\rho_{BC}\otimes\rho_{BC}^{\otimes {(M-1)}})\big)\leq\eps\,.
\end{align}
Here $A'=A^{(M-1)}$ and $B'$ and $C'$ are defined analogously. Like before we can think of $\omega_{A'B'C'}$ as a catalytic resource state that is already locally recovered to start with but has to be kept locally recovered by the randomizing map (cf.~conditional decoupling~\cite{Berta2018}). The {\it non-recoverability cost} denoted by $C_{\mathrm{REC}}(A;B|C)_\rho$ is then defined as the minimal rate $\frac{1}{n}\log M$ needed for $\eps$-recovery-degrading in the limit of asymptotically many copies $\rho_{ABC}^{\otimes n}$ and vanishing error $\eps\to0$. Using again the convex split lemma (Lem.~\ref{lem:convex_split}) and the framework in~\cite{brandao10} for the asymptotic expansion it is straightforward to see that non-recoverability cost is upper bounded by the regularized relative entropy of recovery
\begin{align}\label{eq:upper_bound}
C_{\mathrm{REC}}(A;B|C)_\rho\leq D^{\infty}(A;B|C)_\rho\,.
\end{align}
It would be interesting to understand if this upper bound is also tight. In App.~\ref{app:optimality} we show optimality when restricting the set of allowed coordinated unitary randomizing maps to only include permutations of the $B^M$ systems but arbitrary unitaries on the $A^M$ and $C^M$ systems. Finally, we note that for another well-known measure for tripartite quantum correlations, the {\it conditional quantum mutual information}
\begin{align}
&I(A:B|C)_\rho\notag\\
&:=H(AC)_\rho+H(BC)_\rho-H(ABC)_\rho-H(C)_\rho
\end{align}
we have the typically strict ordering~\cite{Brandao15}
\begin{align}
D^{\infty}(A;B|C)_\rho\leq I(A:B|C)_\rho\,.
\end{align}
Hence, the upper bound in Eq.~\eqref{eq:upper_bound} is in contrast to other recent work about conditional decoupling of quantum information by the authors~\cite{Berta2018} as well as Wakakuwa {\it et al.}~\cite{Wakakuwa17,Wakakuwa17_2}. The fundamental difference is that our final states are locally recovered, i.e.~of the form $(\mathcal{I}_B\otimes\mathcal{R}_{C\to AC})(\rho_{BC})$, but are not themselves (approximately) locally recoverable. In contrast, this is demanded in all of these alternative models.


\paragraph*{Conclusion.}

We have presented a model for catalytic erasure of entanglement in quantum states and showed that the optimal asymptotic rate of noise needed is given by the regularized relative entropy of entanglement. This establishes the disentanglement cost of quantum states as a proper entanglement measure. It would be interesting to work out all the physical consequences of our result -- in the same way as the hypothesis testing interpretation of relative entropy of entanglement~\cite{brandao10} immediately lead to novel insights~\cite{plenio08,plenio10,Gour15}.
We also left open a few questions about extensions to catalytic decoupling models as well as to tripartite quantum correlations in terms of the non-recoverability cost. Finally, our proofs make crucial use of the convex splitting lemma (Lem.~\ref{lem:convex_split}) by Anshu {\it et al.}~\cite{Anshu14} and it would be interesting to better understand all the consequences of this technique in quantum information theory.


\paragraph*{Note.} Our main result Thm.~\ref{thm:distangle} as well as the extension to multi-party entanglement was also derived in the independent work~\cite{Anshu2018}. Moreover, there it is pointed out that the results actually extend to any resource theory that obeys a certain number of natural axioms.


\paragraph*{Acknowledgments.} 

We thank Anurag Anshu, Min-Hsiu Hsieh, and Rahul Jain for discussions. We thank an anonymous referee for pointing out that our results also hold in multi-partite settings. We thank the Institute for Mathematical Sciences at the National University of Singapore for hosting the stimulating workshop Beyond I.I.D.~in Information Theory during which part of this work was done. CM is supported by a NWO VIDI project and acknowledges financial support from the European Research Council (ERC Grant Agreement no. 337603), the Danish Council for Independent Research (Sapere Aude) and VILLUM FONDEN via the QMATH Centre of Excellence (Grant No. 10059).


\bibliographystyle{arxiv_no_month}

\onecolumngrid

\appendix

\section{Non-recoverability cost}\label{app:optimality}

In this appendix, we show that a converse of Eq. (30) also holds: For
coordinated unitary randomising maps where the unitaries $U^i_B$ are
all permutations of the $M$ $B$-systems, the non-recoverability cost
is bounded by the regularized entropy of recovery from below. To this end, for any tripartite state $\sigma_{ABC}$, we define the set
\begin{align}
\mathcal S_\sigma=\big\{\mathcal R_{C\to AC}(\sigma_{BC})|\mathcal R\;\text{local quantum channels}\big\}.
\end{align}
Let now $\rho_{ABC}$ be a fixed quantum state, and let $\omega_{A'B'C'}$ and $\Lambda^M_{AA'BB'CC'}$ define an $\varepsilon$-recovery degrading protocol as described above. We define $\bar L=LL'X_L$ for $L=A,B,C$ and $\alpha_{\bar A\bar B\bar C}=\rho^{\otimes M}\otimes\gamma_{X_AX_BX_C}$. Let further $V_{\bar A\bar B\bar C}=V^a_{\bar A}\otimes V^b_{\bar B}\otimes V^c_{\bar C}$ be a unitary such that
\begin{align}
\tr_{X_AX_BX_C}\left[V\alpha V^{\dagger}\right]=\Lambda^M(\rho\otimes\omega)\;\text{where we have omitted subscripts for brevity.}
\end{align}
Now observe that
\begin{align}
\min_{\sigma\in\mathcal S_\rho}D_{\max}^\varepsilon(\rho_{ABC}\|\sigma_{ABC})=\min_{\sigma\in\hat{\mathcal  S}_{\alpha}}D_{\max}^\varepsilon\Big(\alpha_{\bar A\bar B\bar C}\Big\|\sigma_{\bar A\bar B\bar C}\Big)&=\min_{\sigma\in V\hat{\mathcal  S}_{\alpha}V^\dagger}D_{\max}^\varepsilon\Big(V\alpha_{\bar A\bar B\bar C}V^\dagger\Big\|\sigma_{\bar A\bar B\bar C}\Big)\notag\\
&=\min_{\sigma\in \hat{\mathcal S}_{\alpha}}D_{\max}^\varepsilon\Big(V\alpha_{\bar A\bar B\bar C}V^\dagger\Big\|\sigma_{\bar A\bar B\bar C}\Big)\,.
\end{align}
Here $\hat{\mathcal S}_{\alpha}$ is the set of recovered states for which $\gamma$ is first perfectly recovered and subsequently a recovery map is applied to $CC'$ conditioned on $X_C$. The first equation follows in the same way as in the disentanglement case. The second equation is due to the unitary invariance of the smooth max-relative entropy, and the last equation is due to the fact that
\begin{align}
&\Big(V^a_{\bar A}\otimes V^b_{\bar B}\otimes V^c_{\bar C}\Big)\Big(\mathcal R_{\bar C\to \bar A\bar C}(\rho_{BC}^{\otimes M}\otimes \gamma)\Big)\Big(V^a_{\bar A}\otimes V^b_{\bar B}\otimes V^c_{\bar C}\Big)^{\dagger}\notag\\
&=\Big(V^a_{\bar A}\otimes V^b_{\bar B}\Big)\left(\mathcal R_{\bar C\to \bar A\bar C}\left(\left(\tilde V^c_{\bar B}\right)^{\dagger}\left(\rho_{BC}^{\otimes M}\otimes \gamma\right)\tilde V^c_{\bar B}\right)\right)\Big(V^a_{\bar A}\otimes V^b_{\bar B}\Big)^{\dagger}
\end{align}
for all controlled recovery maps $\mathcal R$. Here $\tilde V^c_{\bar B}$ implements the same controlled permutation on the $B$ systems controlled on $X_B$ instead of the $C$ systems controlled on $X_C$. The above equation holds because of the permutation invariance of $\rho_{BC}^{\otimes M}$. As $\Lambda^M_{AA'BB'CC'}$ and $\omega_{A'B'C'}$ define an $\varepsilon$-recovery degrading protocol, we have that there exists a recovery map $\mathcal R^*_{CC'\to AA'CC'}$ such that
\begin{align}
P\Big(\Lambda^M_{AA'BB'CC'}(\rho_{ABC}\otimes \omega_{A'B'C'}),R^*_{CC'\to AA'CC'}\left(\rho_{BC}^{\otimes M}\right)\Big)\nonumber\le \varepsilon\,.
\end{align}
Now observe that $R^*_{CC'\to AA'CC'}(\rho_{BC}^{\otimes M})\otimes \gamma_{X_AX_BX_C}\in\hat{\mathcal  S}_{\alpha}$, so we can bound
\begin{align}
\min_{\sigma\in \hat{\mathcal  S}_{\alpha}}D_{\max}^\varepsilon\Big(V\alpha_{\bar A\bar B\bar C}V^\dagger\Big\|\sigma_{\bar A\bar B\bar C}\Big)&\le D_{\max}^\varepsilon\Big(V\alpha_{\bar A\bar B\bar C}V^\dagger\Big\|R^*_{CC'\to AA'CC'}(\rho_{BC}^{\otimes M})\otimes \gamma_{X_AX_BX_C}\Big)\notag\\
&\le D_{\max}^0\Big(\beta_{\bar A\bar B\bar C}\Big\|R^*_{CC'\to AA'CC'}(\rho_{BC}^{\otimes M})\otimes \gamma_{X_AX_BX_C}\Big)\,,
\end{align}
where $\beta_{\bar A\bar B\bar C}$ is classical on $X_AX_BX_C$ such that $\left(\Pi_\gamma\right)_{X_AX_BX_C}\beta_{\bar A\bar B\bar C}\left(\Pi_\gamma\right)_{X_AX_BX_C}=\beta_{\bar A\bar B\bar C}$, $R^*_{CC'\to AA'CC'}(\rho_{BC}^{\otimes M})=\mathrm{Tr}_{X_AX_BX_C}\left[\beta_{\bar A\bar B\bar C}\right]$, and $P\left(\beta_{\bar A\bar B\bar C},V(\rho_{ABC}\otimes\omega_{A'B'C'}\otimes\gamma_{X_AX_BX_C})V^{\dagger}\right)\le\varepsilon$. The existence of such a state $\beta$ follows again by Uhlmanns theorem. Applying the operator inequality Eq.~\eqref{eq:matrixineq} in the same way as in the disentanglement case finishes the proof. \qed


\end{document}